\documentclass[aps, prb, amsmath, amssymb, preprint,
               groupedaddress, showpacs]{revtex4}

\usepackage{graphicx}% Include figure files
\usepackage{bm}% bold math
\usepackage[dvips]{color}
\usepackage{subfigure}

\begin{document}
    \DeclareGraphicsExtensions{.jpg,.png,.mps,.pdf,.eps}

\title{Charge ordering of magnetic monopoles in triangular spin ice patterns}
 \author{A. Schumann$^1$, B. Sothmann$^2$, P. Szary$^1$, and H. Zabel$^1$}

 \affiliation{$^{2}$Institut f\"{u}r Experimentalphysik/Festk\"{o}rperphysik,
Ruhr-Universit\"{a}t
Bochum, 44780 Bochum, Germany\\
$^{2}$Theoretische Physik, Universit\"{a}t Duisburg-Essen and
CeNIDE, 47048 Duisburg, Germany}

\begin{abstract}
Artificial spin ice offers the possibility to investigate a
variety of dipolar orderings, spin frustrations and ground states.
However, the most fascinating aspect is the realization that
magnetic charge order can be established without spin order. We
have investigated magnetic dipoles arranged on a honeycomb lattice
as a function of applied field, using magnetic force microscopy.
For the easy direction with the field parallel to one of the three
dipole sublattices we observe at coercivity a maximum of spin
frustration and simultaneously a maximum of charge order of
magnetic monopoles with alternating charges $\pm$ 3.

\end{abstract}


\maketitle

In natural spin ice, magnetic ions form a network of
corner-sharing tetrahedra such as holmium ions in holmium
titanate\cite{Ramirez1999}. Pairs of their magnetic moments point
either in or out of the tetrahedra, which is then termed the "two
in - two out" rule, following water ice, where on the average two
hydrogen atoms point towards an oxygen ion and two point away. The
two in - two out configuration is favored by the magnetic
dipole-dipole interaction. Other configurations also occur
frequently, such as three in and one out, or four in and none out.
All these configurations carry a certain amount of frustration,
but the best compromise still is the two in - two out rule,
resulting in long range order. Modern lithographic techniques
enable the fabrication of magnetic dipoles arranged on lattices
with different symmetries, mimicking two dimensional projections
of spin ice. Artificial square and triangular spin ice patterns
are of particular interest in current research\cite
{Wang2006,Tanaka2006,Qi2008,Zabel2009}, because they provide a
laboratory for the analysis of order and excitations in frustrated
planar lattices with dipolar interaction\cite{Moessner2006}.

\begin{figure}[h]
\includegraphics[width=0.5\textwidth]{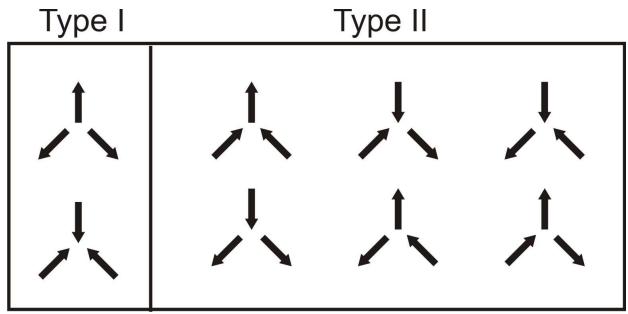} \caption{Possible
configurations of magnetic dipoles on a honeycomb lattice.
}\label{TypI-TypII-a}
\end{figure}

Here we consider dipolar arrays placed on a honeycomb lattice,
where each vertex consists of three equivalent dipoles enclosing
an angle of 120$^{\circ}$.  The honeycomb lattice exhibits a
fascinating variety and complexity of configurations and at the
same time is highly frustrated. For each vertex there are a total
of 2$^3$=8 possible configurations. In analogy to the square spin
ice, the spin ice rule for the honeycomb lattice is fulfilled if
two dipoles point in and one points out, or vice versa. There are
six configurations, which fulfill this rule (type II). The
remaining two spin configurations (type I) with all three dipoles
pointing either in or out, violate the spin ice rule. At the same
time, the type I and II configurations define a certain value of
"magnetic charges" (see Fig.~\ref{TypI-TypII-a}). Assigning a
charge +1 for dipoles pointing in and -1 for dipoles pointing out,
type I vertices carry a charge of $\pm$ 3, and type II
configurations a charge of $\pm$ 1. Vertices with charge $\pm$ 3
have no effective magnetic dipole and can be described as magnetic
monopoles. It was recently pointed out that irrespective of the
configurations adapted by an artificial spin ice pattern, the
total charge of a pattern must be zero\cite{Moessner2006}. There
is a another important prediction concerning spin ice
configurations: charge order may occur independent of spin
order\cite{Moeller2009}. In fact, in many systems charge order is
found on different temperature scales than spin order. Examples
are layered nickelates\cite{Giovannetti2009} or
cobaltates\cite{Lang2008}. In artificial spin ice patterns we
cannot probe the temperature dependence of the ordering as thermal
fluctuations are essentially suppressed by the large shape
anisotropy. Instead we present field dependent studies, where spin
order is induced in saturation and apparently maximum disorder is
found at coercivity. However, charge order of magnetic monopoles
may still be present.

The honeycomb lattice has been studied theoretically by several
authors\cite{Moessner2006,Moeller2009,Wills2002} and was realized
experimentally by lithographic means and investigated via magnetic
force microscopy\cite{Tanaka2006}, Lorentz microscopy\cite{Qi2008}
and photoemission electron microscopy\cite{Mengotti2008}. In the
first two cases the dipoles were connected to each other, yielding
the strong interaction limit, while in the third case only single
up to triple honeycomb rings were investigated, excluding a
statistical analysis. Here we present experimental realizations of
magnetic dipoles arranged on a honeycomb lattice and we discuss
the remanent state as well as the magnetization reversal in an
external field. The reversal strongly depends on the field
direction with respect to the main symmetry axes of the pattern,
which are either [10] or [11]. In the [10] direction (hard axis),
one dipole set of the three sublattices is perpendicular to the
field direction and the other two are inclined at an angle of
30$^{\circ}$. For the [11] direction (easy axis), one dipole set
is parallel to the field direction, whereas the other two are
inclined at an angle of  60$^{\circ}$. We mainly focus on the [11]
orientation and find that at coercivity the magnetization reversal
passes through a highly charge ordered  state.

\begin{figure}
\includegraphics[width=0.8\textwidth]{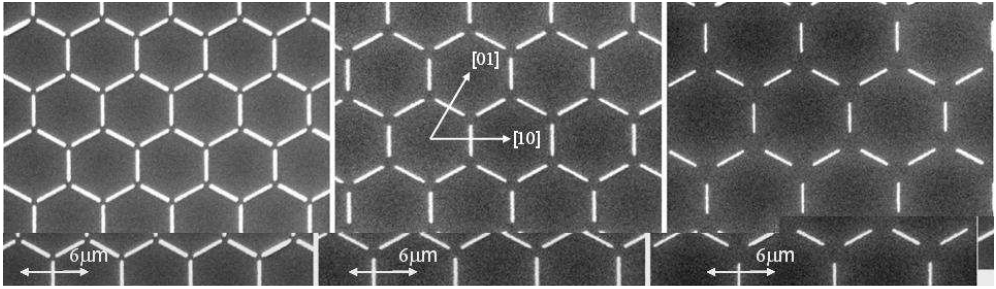}\caption{SEM-image of three honeycomb lattices
with different inter island spacing. Left: 0.4 $\mu$m, Middle: 0.8
$\mu$ m, Right: 1.7 $\mu$ m. Red bars indicate the scale and red
arrows show the basis vectors of the honeycomb lattice.
}\label{SEM-alle}
\end{figure}

We deposited a 20 nm thick polycrystalline Fe film on a silicon
substrate with a 5 nm thick Ta seed layer capped by a 2 nm thick
Al$_2$O$_3$ layer for oxidation protection. The honeycomb patterns
were prepared by means of e-beam lithography and ion beam etching,
following procedures described elsewhere\cite{Remhof2008}. The
final patterns consist of Fe-bars with dimensions length, width,
and thickness of 3 $\mu$m, 0.3 $\mu$m and 20 nm, respectively.
Here we discuss three honeycomb patterns with inter-island
distances of 0.4 $\mu$m, 0.8 $\mu$m and 1.7 $\mu$m, SEM images of
the patterns are shown in Fig.~\ref{SEM-alle}. The dipole
configurations were imaged by a magnetic force microscope (MFM) at
room temperature equipped with tunable in-plane magnetic field up
to 1000 Oe and a rotation stage for aligning the pattern with
respect to the external field. In the MFM images each
ferromagnetic island has at the ends a bright and a dark spot due
to the emanating stray fields. This contrast confirms that for the
aspect ratio chosen the islands are dominantly in a single domain
state.

First we demagnetized the samples by driving the pattern  through
minor loops, meaning that we carefully decreased the applied
magnetic field from well above saturation (1000 Oe) in small steps
of 10 Oe while changing the direction of the magnetic field with
each step \cite{PhysRevLKe,JAPWang}. Then we recorded MFM-images
and analyzed the magnetic charge and the total magnetization of
the scanned area by image processing. For evaluating the
magnetization we assigned dipoles aligned parallel to the applied
magnetic field with the normalized moment of $\pm$ 1, all others
contribute a normalized moment of
$\textrm{cos}(\pm60^{\circ},\pm120^{\circ})$ = $\pm$ 1/2. By
adding up all moments we derive a magnetization for each field
value and thus a digital magnetic hysteresis curve.

\begin{figure}[h!]
\subfigure[]{
\includegraphics[scale=1.0]{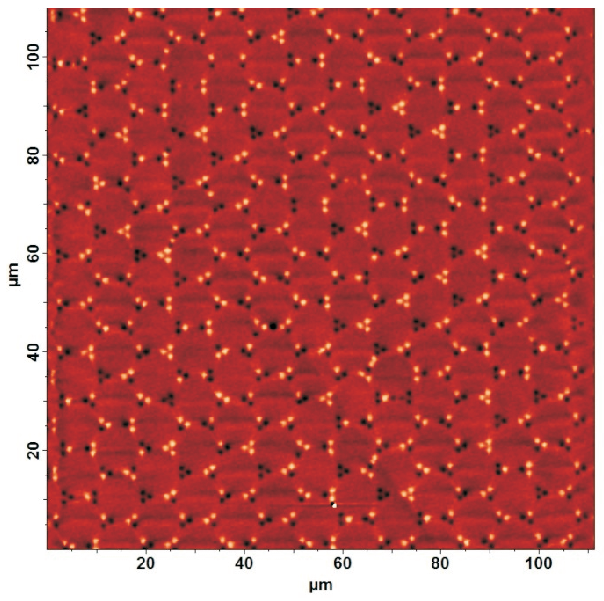}}
\subfigure[]{
\includegraphics[scale=1.0]{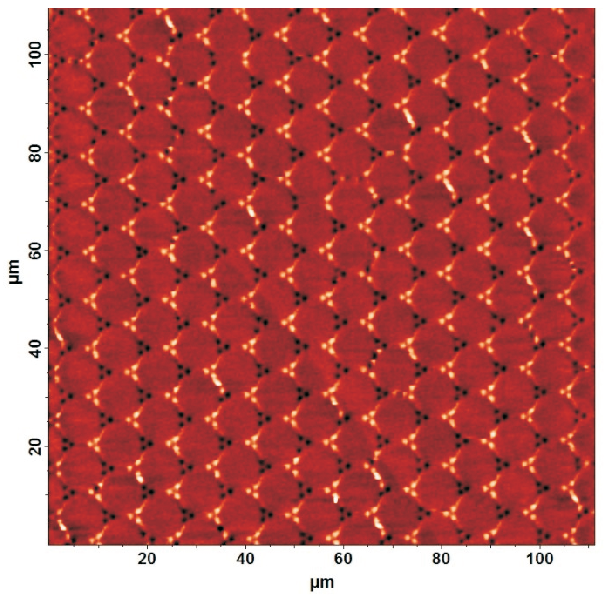}}
\caption{\label{MFM}(a): MFM-image  of a demagnetized honeycomb
lattice with inter-island spacing of 0.8 $\mu$m. (b): MFM-image of
the honeycomb pattern with an inter-island distance of 0.8 $\mu$m
taken at H$_{ext}$ = -500 Oe. Note the high order of alternating
$\pm$3 magnetic charges at the vertices.}
\end{figure}

For the [11] orientation (one sublattice aligned parallel to the
applied magnetic field) and for large inter island distances we
observe a nearly uncorrelated demagnetized state with a frequency
of type I and type II vertices as expected for a random
distribution. At the same time the demagnetized state fulfills the
condition of charge neutrality. A representative example for the
demagnetized pattern with a dipole separation of 0.8 $\mu$m is
shown in Fig.~\ref{MFM}(a).

\begin{figure}[h!]
\begin{center}
\subfigure[]{
\includegraphics[scale=1.0]{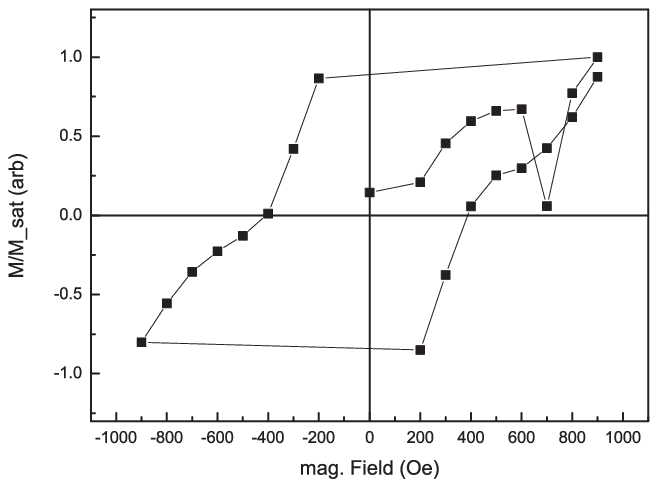}}
\subfigure[]{
\includegraphics[scale=1.0]{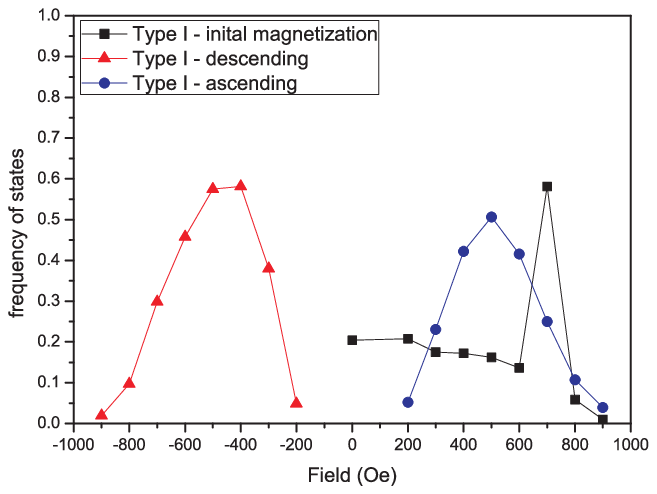}}
\caption{\label{Hys}(a): Digital hysteresis loop  for the magnetic
field applied parallel to the [11] direction. (b): Frequency  of
the type I-state as a function of external field. }
\end{center}
\end{figure}

We found that in the demagnetized state the charge and the
magnetization is close to zero, which means that our
demagnetization protocol was successful. Small deviations can be
attributed to counting errors in the analysis software and/or
local defects in the dipolar array. Upon increasing the external
field parallel to the [11] direction, the magnetization first
increases, drops again and finally reaches the saturation value.
Fig.~\ref{Hys}(a) reproduces the initial magnetization after the
demagnetization procedure has been applied and a full loop for the
honeycomb pattern with a dipole separation of 0.8 $\mu$m. Each
point in this hysteresis is the result of a numerical evaluation
of all dipoles in the scanned region of the pattern, as described
before. Figure \ref{Hys}(b) shows the frequency of the type I
state for this pattern and for the [11] field orientation. We
notice that for the [11]  orientation the frequency of type I
vertices dramatically increases at coercivity in the descending as
well as in the ascending branch of the hysteresis. The frequency
of type I reaches values as high as 70\%. This is rather
surprising as type I vertices with a charge $\pm$3 violate the
spin ice rule and present local maxima in the potential landscape
for the magnetization reversal. At the same time we observe at
coercivity a very high degree of charge order of magnetic
monopoles with alternating charge $\pm$3. Figure \ref{MFM}(b)
shows the 0.8 $\mu m$ pattern at the field H$_{ext}$ = -500 Oe
descending from saturation. We infer from this image that
coercivity in the honeycomb lattice is characterized not only by
zero magnetization but also by the highest possible spin
frustration with nearly complete charge order of magnetic
monopoles.

The magnetization reversal through the charge ordered state can be
understood by considering the coercivity fields of all three
sublattices. The horizontal dipoles parallel to the external field
direction switch first in a reversal field, followed by switching
of the 60$^{\circ}$ inclined dipoles. As switching proceeds via a
domain wall process, the effective switching field acting on the
inclined dipoles is only half the value compared to that acting on
the horizontal dipoles. Therefore between the first and second
coercive field charge order may occur. We may speculate that in
this region the frequency of type I states could increase to 100\%
and the charge order could be complete, if a certain distribution
of switching fields and local defects in the pattern could be
avoided. In this scenario maximum spin frustration coexists with
complete charge order of magnetic monopoles.

The magnetization reversal with the field applied parallel to the
[10] direction follows a completely different path. The dipoles
oriented perpendicular to the field direction do not switch at all
in the field range applied here. They remain in a random
orientation as established after the demagnetization procedure.
The other dipoles inclined at an angle of 30$^{\circ}$ against the
field direction switch more or less simultaneously. Thus neither
magnetic order nor charge order can be established in this
orientation. However, even if the perpendicular dipoles were first
aligned parallel, charge order, as observed in the [11] direction
at coercivity, would never prevail.

In summary, we have fabricated an artificial spin ice structure by
arranging magnetic dipoles on a honeycomb lattice using e-beam
lithography. We have studied the magnetization reversal of the
honeycomb pattern for field directions parallel to the main
symmetry directions of the triangular lattice, [10] and [11]. For
the magnetic field applied parallel to the [11] direction we find
at coercivity a highly ordered state of magnetic monopoles with
alternating charges $\pm$ 3. Perfect order is only hindered by
local defects and by a distribution of switching fields. In
contrast, for fields applied along the [10] direction, a charge
ordered state can not be established.

The authors are grateful for technical support by Peter  Stauche.
We would like to thank the Deutsche Forschungsgemeinschaft for
financial support of this work within the SFB 491.

%\end{multicols}{2}

\cleardoublepage

\end{document}